# Phase Fluctuations and the Absence of Topological Defects in Photo-excited Charge Ordered Nickelate

W. S. Lee[1†]*, Y. D. Chuang[2†], R. G. Moore[1], Y. Zhu[7], L. Patthey[1,3], M. Trigo[1,5], D. H. Lu[4], P. S. Kirchmann[1], , O. Krupin[6,9], M. Yi[1], M. Langner[7], N. Huse[7,10], J. S. Robinson[7], Y. Chen[1], S. Y. Zhou[2,7], G. Coslovich[7], B. Huber[7], D. A. Reis[1,5], R. A. Kaindl[7], R. W. Schoenlein[2,7], D. Doering[8], P. Denes[8], W. F. Schlotter[9], J. J. Turner[9], S. L. Johnson[3], M. Först[10], T. Sasagawa[11], Y. F. Kung[1], A. P. Sorini[1], A. F. Kemper[1], B. Moritz[1], T. P. Devereaux[1], D.-H. Lee[12], Z. X. Shen[1]*, and Z. Hussain[2]

[1]*Stanford Institute for Materials and Energy Science, SLAC National Accelerator Laboratory and Stanford University, Menlo Park, CA 94025, USA*

[2]*Advanced Light Source, Lawrence Berkeley National Laboratory, Berkeley, CA 94720, USA*

[3]*Swiss Light Source, Paul Scherrer Institut, CH-5232 Villigen-PSI, Switzerland*

[4]*Stanford Synchrotron Radiation Lightsource, SLAC National Accelerator Laboratory, Menlo Park, CA, 94025, USA*

[5]*Stanford PULSE Institute for Ultrafast Energy Science, SLAC National Accelerator Laboratory, Menlo Park, CA 94025, USA*

[6]*European XFEL, Hamburg, Germany*

[7]*Materials Science Division, Lawrence Berkeley National Laboratory, Berkeley, CA 94720, USA*

[8]*Engineering Division, Lawrence Berkeley National Laboratory, Berkeley, CA 94720, USA*

[9]*Linac Coherent Light Source, SLAC National Accelerator Laboratory, Menlo Park, CA 94720, USA*

[10]*Max-Planck Research Group for Structural Dynamics, University of Hamburg, CFEL, Germany*

[11]*Materials and Structures Laboratory, Tokyo Institute of Technology, Kanagawa 226-8503, Japan*

[12]*Department of Physics, University of California at Berkeley, Berkeley, CA 94720, USA*

[†]Y. D. Chuang and W. S. Lee led the project and have equal contribution to this work.

*To whom correspondence should be addressed; e-mail: leews@stanford.edu and zxshen@stanford.edu.*

**The dynamics of an order parameter's amplitude and phase determines the collective behaviour of novel states emerged in complex materials[1]. Time- and momentum-resolved pump-probe spectroscopy, by virtue of its ability to measure material properties at atomic and electronic time scales[2,3,4,5,6,7,8,9] and create excited states not accessible by the conventional means can decouple entangled degrees of freedom by visualizing their corresponding dynamics in the time domain. Here, combining time-resolved femotosecond optical and resonant x-ray diffraction measurements on striped $La_{1.75}Sr_{0.25}NiO_4$ , we reveal unforeseen photo-induced phase fluctuations of the charge order parameter. Such fluctuations preserve long-range order without creating topological defects, unlike thermal phase fluctuations near the critical temperature in equilibrium[10]. Importantly, relaxation of the phase fluctuations are found to be an order of magnitude slower than that of the order parameter's amplitude fluctuations, and thus limit charge order recovery. This discovery of new aspect to phase fluctuation provides more holistic view for the importance of phase in ordering phenomena of quantum matter.**

In striped nickelates, charge (CO) and spin order (SO) coexist at low temperatures, where doped charge carriers form one dimensional charge density waves, and spins align in spin order with half the CO period (Fig. 1a).[11,12,13,14,15,16] Similar stripe phases also have been found in doped cuprates[17,18], where they are thought to compete against high temperature superconductivity. In the present work, both time-resolved optical reflectivity and resonant x-ray diffraction[19] (RXD) were used to reveal the dynamics of the CO order parameter. In particular, RXD at the Ni $L$-edge primarily measures the Bragg scattering from the valence electron modulation of Ni ions at the CO wave vector. The intensity is proportional to the CO order parameter's amplitude and a Debye-Waller like factor due to vibrations of the electronic CO[20,21,22] (i.e. the phase fluctuations). In addition, the correlation length of CO can be inferred from the width of the CO diffraction peak. Therefore, the recent availability of the femtosecond x-ray free electron laser (FEL) at Linac Coherent Light Source[23] makes time-resolved RXD near the Ni $L_3$- edge an ideal probe (Fig. 1b) to measure the dynamics of electronic CO order parameters with femtosecond resolution. A single bright x-ray FEL pulse can generate analyzable diffraction patterns for the CO (Fig. 1c), allowing pulse-by-pulse data acquisition for correcting the x-ray FEL timing and intensity jitter in the data analysis. The overall temporal resolution is ~ 0.4 ps limited by the synchronization between the pump laser and x-ray FEL pulse[24]. Details of the experimental setup are described in the Supplementary Information.

Fig. 2a shows the early dynamics of the integrated intensity of the CO diffraction peak for two different pump excitation densities. Following the optical pump pulse at time zero, the intensity drop sharply (region I), indicating a photo-induced suppression of CO. After reaching a minimum, the intensity recovers toward its original value (region II). Fig. 2b displays the evolution of the CO diffraction peak for a given excitation density. Within our experimental accuracy, no variation of the peak position can be resolved in either region I or II, indicating that the period of CO essentially remains unchanged in the photo-excited transient state. Importantly, the width of the CO peak shows no sign of broadening in the investigated time range, suggesting that the coherence length does not shorten due to photo-excitation despite a significant reduction in the peak intensity. This is in sharp contrast to thermal evolution of the CO diffraction peak near the equilibrium phase transition temperature where the dramatic decrease of the peak intensity is accompanied by a width broadening[12,13,14] ( Fig. 2c). Thus, in the photo-induced non-equilibrium state, CO peak intensity can vary dramatically without changing the CO length scales.

The time scales exhibited in the recovery reveal further information about the CO dynamics. As demonstrated in Fig. 3a, the recovery of CO peak intensity (up to $\Delta t = 25$ ps) is characterized by two recovery processes with distinct time scales, $\tau_{co}^{fast}$ ~ 2 ps and $\tau_{co}^{slow}$ ~ 60 ps, which differ by more than one order of magnitude (see Supplementary Materials for fitting procedures). Furthermore, the component of the slower dynamics is significant, which limits the recovery of the CO peak intensity. To gain further insight into the origin of these two time scales, it is informative to compare RXD data with optical reflectivity data, which is sensitive to the local charge dynamics[5,6]. The recovery of photo-induced reflectivity change Δ$R/R$ (Fig. 3b) is similarly characterized by two recovery processes: a faster and dominant component of $\tau_{\Delta R/R}^{fast}$~ 0.5 ps, plus a slower recovery of $\tau_{\Delta R/R}^{slow}$ ~2ps. Comparing the time scales deduced from the two techniques, we found $\tau_{co}^{fast}$ is similar to $\tau_{\Delta R/R}^{slow}$; however, the dynamics corresponding to $\tau_{co}^{slow}$ and $\tau_{\Delta R/R}^{fast}$ are not evident in the optical reflectivity and RXD data, respectively. These observations are further confirmed by the pump excitation dependence summarized in Fig. 3c. Clearly, three different dynamics are involved, and both $\tau_{co}^{fast}$ and $\tau_{\Delta R/R}^{slow}$ are similar within the range of investigated pump excitation density. Also, $\tau_{co}^{slow}$ increases more pronounced than other time scales as the pump excitation density increases, implying a different origin than that of $\tau_{co}^{fast}$ for long range CO recovery.

We note that Eichberger *et al.*[4] have reported a similar observation by comparing recovery time scales of optical reflectivity to those of femtosecond electron diffraction (FED) in TaS$_2$. Since FED is only sensitive to atomic positions, the distinct time scales have been attributed to the distinction between electronic and lattice order parameters. In contrast, our time-resolved RXD measurements provide hereto unforeseen behavior of the *electronic* order parameter dynamics, as the CO diffraction peak at the Ni *L$_3$*-edge in striped nickelates comes primarily from light scattered by the Ni valence electron density modulation[15]. Therefore, the recovery time scales deduced from the optical reflectivity and RXD are both of *electronic* origin, but correspond to three different dynamics in the photo-induced non-equilibrium states. First, since RXD is directly sensitive to the long range CO, the absence of the fastest dynamics, $\tau_{\Delta R/R}^{fast}$ ~ 0.5 ps, suggests that this kinetics is linked to microscopic relaxation processes of non-equilibrium electronic states and is not directly linked to the restoration of long ranged CO. Second, the similar time scales between $\tau_{co}^{fast}$ and $\tau_{\Delta R/R}^{slow}$ suggests that they likely share a common origin. As this dynamics is seen in both the charge distribution channel *(ΔR/R)* and the long range order channel (XRD), it is reasonable to attribute this dynamics to the recovery of the magnitude of the charge density modulation, i.e. the CO order parameter's amplitude.

Most interestingly, the recovery of the order parameter's amplitude does not occur simultaneously with the recovery of the CO diffraction peak intensity and indicates the importance of phase fluctuations in determining the recovery time of the CO itself. Phase fluctuations could be viewed as vibrations of the electronic lattice (i.e. the periodic CO-related charge density modulation), analogous to vibrations of the crystal lattice. Although the exact functional form is somewhat debated, it has been demonstrated for charge density wave systems that fluctuations of the order parameter's phase can qualitatively give rise to a Debye-Waller suppression of the CO diffraction peaks in addition to the lattice Debye-Waller factor[20, 21, 22]. As the RXD signal is mostly sensitive to electronic CO scattering; only the phase Deybe-Waller factor will be exclusively captured in our time-resolved XRD data. In accordance with our results, the slower component of the CO peak intensity recovery ($\tau_{co}^{slow}$) is most likely dominated by diminishing phase fluctuations of the CO order parameter. We note that phase excitations can be produced, because the phase excitations are often gapless or acquire a small gap due to pinning by the lattice or defects, and hence, require little energy to excite. In addition, since the phase Debye-Waller factor exponentially suppresses the CO peak intensity scaling with the square of average phase fluctuation, small amounts of phase fluctuation can significantly suppress the CO peak intensity[20].

Therefore, $\tau_{co}^{slow}$ is a direct measure of how fast energy can be transferred out of phase fluctuations and contains information about how strongly phase excitations (i.e. phasons) couple to the lattice (i.e. phonons). In the limit of zero excitation density, the phase fluctuation magnitude is infinitesimally small, thus $\tau_{co}^{slow}$ can be thought to represent the intrinsic equilibration phason-phonon coupling time scale, $\tau_{\phi\text{-ph}}$. By extrapolating to zero excitation density in Fig. 3b, we estimate $\tau_{\phi\text{-ph}}$ to be approximately 15 ps. We note that this is a lower bound of the phason-phonon relaxation time scale. Pinning effects introduced by disorder that limits the length scale of CO may also provide another mechanism for phason damping. High voltage resistivity measurements suggest that these effects are non-negligible[25]. Nevertheless, $\tau_{\phi\text{-ph}}$ is more than one order of magnitude slower than the electron-phonon relaxation time, which is estimated to be ~ 0.4 ps by extrapolating $\tau_{\Delta R/R}^{fast}$ to zero excitation density. This suggests that phason-lattice coupling is still much weaker than electron-phonon coupling.

The dynamics of CO in the photo-excited transient state are summarized in Fig. 4a. At the earliest times in region I, the electronic energy abruptly increases due to the absorption of photons from the pump, causing the CO amplitude (Δ) suppression and generating phase (ϕ) fluctuations. The excited electrons first decay rapidly, shedding energy e.g. into lattice relaxation (~ 0.5 ps), and the CO amplitude recovers on a time scale of ~ 2 ps. In region II, the amplitude of the CO order parameter has mostly recovered, but the phase continues to fluctuate, accounting for most of the remaining excited electronic energy. It is found that the time scale for equilibration between the phase fluctuation and the lattice is long ( > 15 ps), hence limiting the recovery. In addition, CO remains long-ranged without changing its period in either region I or II; this indicates that while the photo-induced transient state involves the amplitude and phase fluctuations, topological defects are not created as it requires reconfiguring a large number of spins and charges at a large energy cost. This is in stark contrast to the physics near the equilibrium phase transition temperature (Fig. 4b), where low energy topological defects[10] reduce the correlation length and the amplitude of the CO order parameter.

Finally, we remark that the physics described in our study should be generally applicable to the recovery of complex order parameters in other systems, as long as the pump excitation density is not strong enough to excite defects in the phase ordering and when the phase modes are soft and couple weakly to the lattice. Photo-induced phase fluctuations may be crucial to understanding the mechanism of photo-induced superconductivity in the striped cuprates[2].

**Acknowledgement**s This research was supported by the U.S. Department of Energy, Office of Basic Energy Sciences, Division of Materials Sciences and Engineering under Contract No. DE-AC02-76SF00515, SLAC National Accelerator Laboratory (SLAC), Stanford Institute for Materials and Energy Science (W.S.L., R. M., L.P., M.T, Y.C., D.A.R, Y.F.K., A.P.S., A.F.K., B.M, T.P.D., Z.X.S), SLAC Stanford Synchrotron Radiation Lightsource (D.H.L.), SLAC Stanford PULSE Institute (M.T., D.A.R.) and under contract number DE-AC02-05CH11231 Lawrence Berkeley National Laboratory (LBNL) Advanced Light Source (Y.D.C., Z.H.), LBNL Materials Sciences Division (M.L., J.R., Y.Z., S.Z., G.C., B.H. R.A.K., R.W.S.), LBNL Chemical Science Division (N. H.), and LBNL Engineering Division (D.D., P.D.). P.S.K acknowledges support by the Alexander-von-Humboldt Foundation through a Feodor-Lynen scholarship. Y.F.K. was supported by the Department of Defense (DoD) through the National Defense Science & Engineering Graduate Fellowship (NDSEG) Program. D.H.L acknowledges the support by the DOE grant number DE-AC02-05CH11231. The SXR Instrument at LCLS is funded by a consortium whose membership includes LCLS, Stanford University - SIMES, LBNL, University of Hamburg through the BMBF priority program FSP 301, and the Center for Free Electron Laser Science (CFEL).

**Author Contribution:** W.S.L. and Y.D.C. conceived and led the project, and contributed equally to this work. T.S. grew the single crystals for the experiments. Y.D.C. designed the RSXS endstation. For the experiment at LCLS, Y.D.C., W.S.L., R.G.M., L.P., M.T., D.H.L., P.S.K., M.Y., O.K., W.F.S., J.J.T., D.D., P.D. setup RSXS endstation and M.T. M.L., N.H., J.S.R., Z.Y., S.Y.C., D.A.R., R.A.K., R.W.S., S.L.J. setup the pump laser. Y.D.C., W.S.L., R.G.M., L.P., M.T., D.H.L., P.S.K., M.Y., O.K., W.F.S., J.J.T., D.D., P.D., Y.C., M.L., N.H., J.S.R., Z.Y., S.Y.C., D.A.R., R.A.K., R.W.S., M.F., Z.X.S., Z.H. participated the experiment at the LCLS. Y.Z., G.C., B.H. performed optical reflectivity measurement. P.S.K. measured the sample reflectivity. R.G.M. and M.Y. composed computer program for extracting images from data taken at the LCLS. R.G.M. and W.S.L. analyzed the data. D.H.L., Y.F.K., A.P.S., A.F.K., B.M., T.P.D. provided theoretical insight into the data interpretation and analysis.

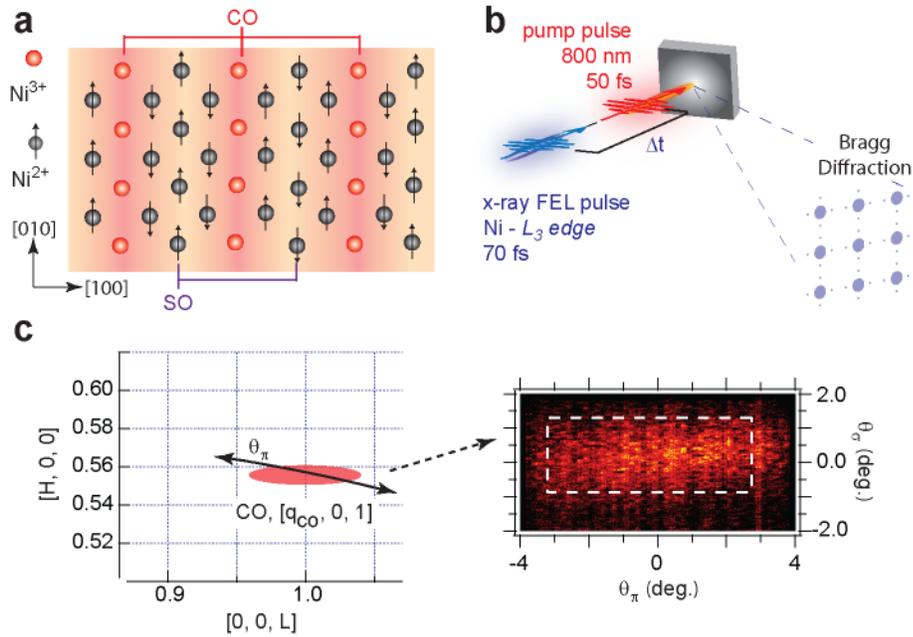

Fig. 1

**Fig. 1 Striped nickelate.** (a) A cartoon of the stripe state in doped nickelate with doping level x=0.25. In this simplified ionic cartoon picture, doped holes mostly reside on $Ni^{3+}$ ions, while the remaining $Ni^{2+}$ ions are at high spin state (S = 1). (b) A sketch of the optical pump and RXD resonant diffraction probe experiment. At time zero ($\Delta t = 0$), the sample was pumped by an optical laser pulse (800 nm, 50 fs); an x-ray free electron laser (FEL) probe pulse with a temporal duration less than 70 fs was introduced at a time delay $\Delta t$ to map out the time evolution of the CO resonant diffraction peaks. (c) Reciprocal space diagram for the locations of the CO diffraction peaks. $q_{co} \sim 0.554$ for the measured sample. Also shown are the CO diffraction peaks captured by a fast CCD detector[16]. The diffraction peak images are produced by a single x-ray FEL pulse tuned to 851 eV, corresponding to the Ni $L_3$-edge ($2p - 3d$ transition). The horizontal axis ($\theta_\pi$) of the image is parallel to the scattering plane, corresponding to the cut (black solid curve) through the diffraction peak in the [H, 0, L] plane. The vertical axis $\theta_\sigma$ is perpendicular to the scattering plane, corresponding to the [0, K, 0] direction.

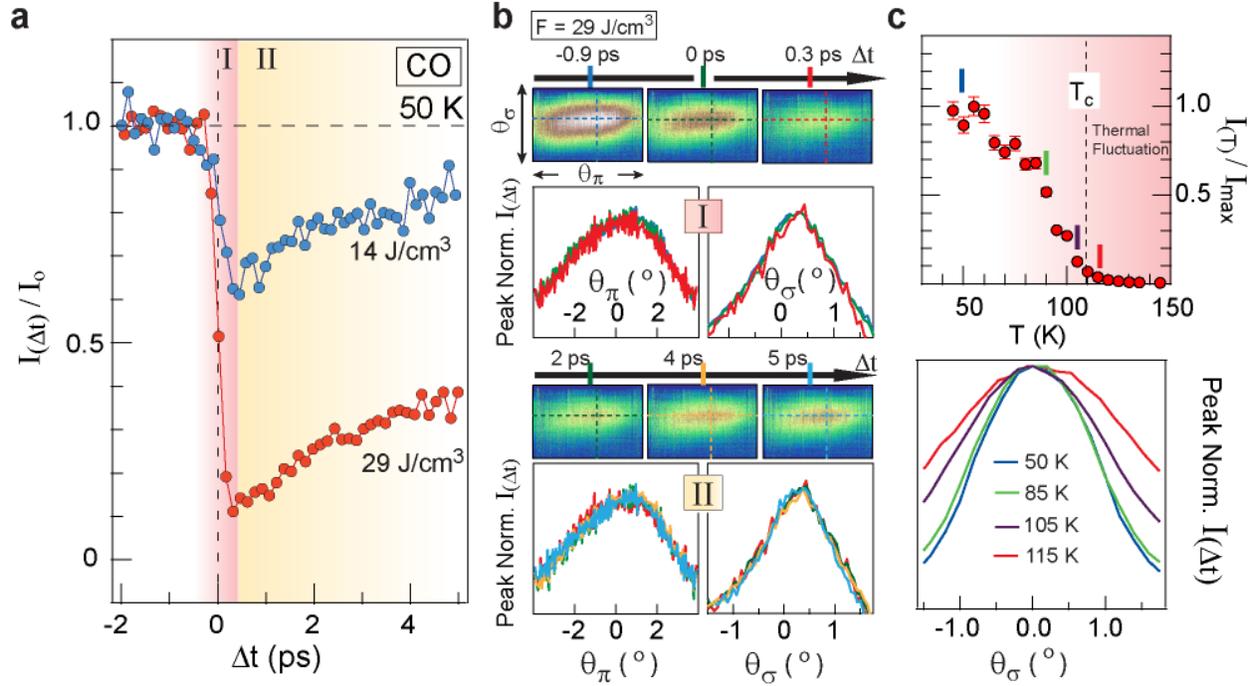

Fig. 2

**Fig. 2 Early dynamics of the charge order.** (a) Integrated intensity within the region indicated by the white-box in Fig. 1c plotted within the first 5 ps after time zero. The pump excitation density is indicated by associated labels. (b) Evolution of the CO peak at the excitation density of 29 J/cm$^3$ in regions I (upper row) and II (lower row). Images at representative time delays were obtained from an average of approximately 30 – 40 x-ray FEL pulses. Below the images, peak-intensity-normalized line-cuts along the dashed lines are plotted with different colors corresponding to color-coded representative time delays. (c) Temperature dependence of the CO resonant diffraction peak measured in thermal equilibrium conditions at a synchrotron light source. The peak intensities are plotted in the upper panel. Lower panel displays peak-intensity-normalized line-cuts along $\theta_\sigma$ at representative temperatures (indicated by color ticks in the upper panel).

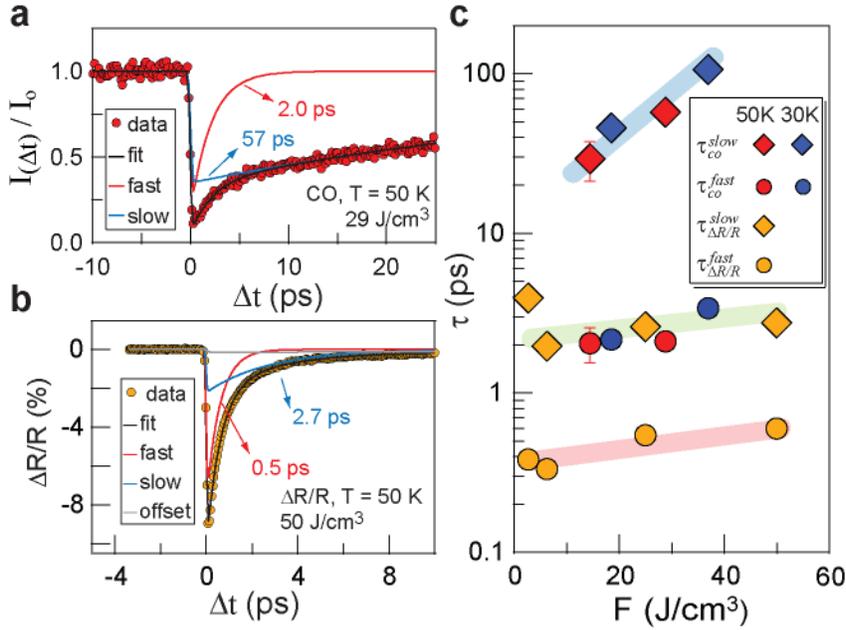

Fig. 3

**Fig. 3 Comparison of recovery time scales.** (a) Time dependent integrated intensity of the CO diffraction peak, which were fitted to two exponential functions. The fitted time scales are indicated. (b) Time dependent reflectivity change $\Delta R/R$ at a wavelength of 800 nm. The time traces were fitted to a fast and slow exponential recovery in addition to a constant background, representing a long-living meta-stable state in the local charge channel. The fitted time scales are indicated. (c) Fast and slow recovery time scales deduced from the integrated intensity of CO ($\tau_{CO}$) and optical $\Delta R/R$ data ($\tau_{\Delta R/R}$). No difference can be discerned between the dynamics measured at 30K and 50 K. Temporal resolution were taken into account as a Gaussian convolution in the fitting processes.

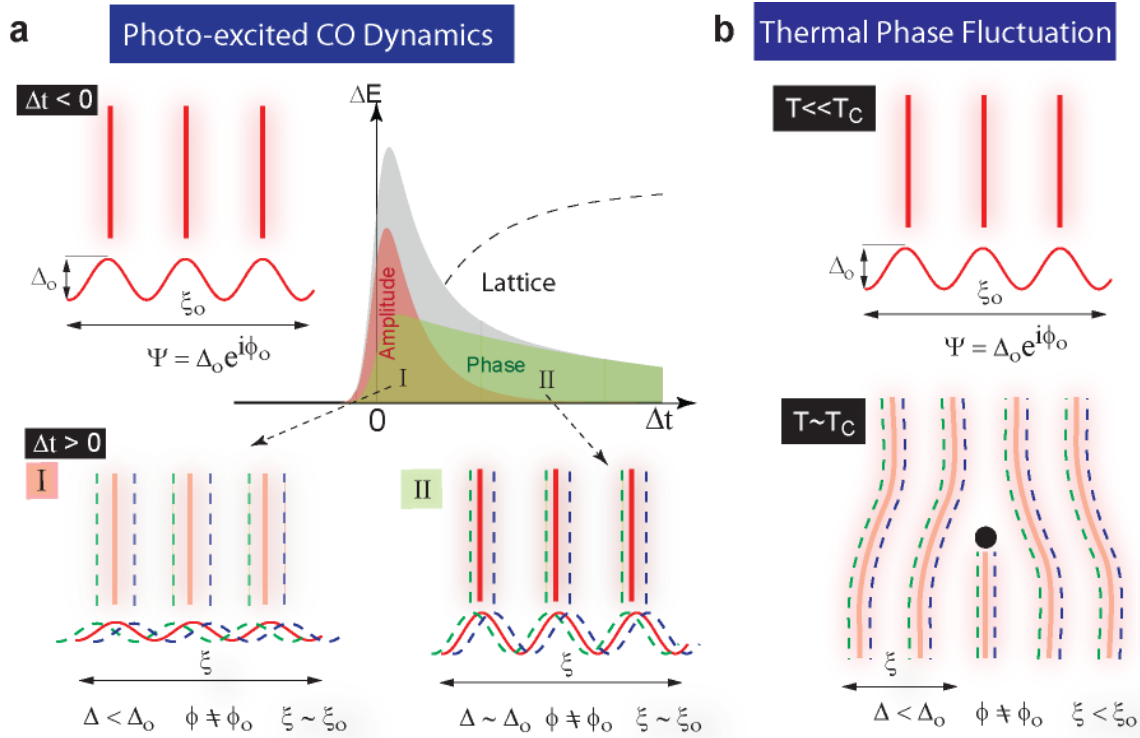

Fig. 4

**Fig. 4 Illustration of CO order parameter dynamics.** (a) The photo-excited CO dynamics. ΔE is the energy difference with respect to the ground state energy. The grey shaded areas represent the total ΔE in the long-ranged CO channel. The dissipation of ΔE through the coupling to the lattice results in the recovery of the amplitude and phase of the CO order parameter. The order parameter completely recovers when the absorbed energy eventually dissipates entirely to sample environment. Sketches of order parameters at three representative temporal regions are also shown. In each sketch, the red stripes illustrate the CO (i.e. the charge stripe) in real space whose order parameter is symbolically drawn below as the red sinusoidal wave. The brightness of the stripe also represents the amplitude of the order parameter. The dashed sinusoidal waves illustrate the fluctuations of the phase of the CO order parameter. In these sketches, Δ, ϕ, and ξ respectively represent the amplitude, phase, and the phase-phase correlation length of the CO order parameter. (b) Sketches for CO evolution in thermal equilibrium. Near the phase transition temperature $T_C$, in addition to the Gaussian phase fluctuation, large phase fluctuations also occur

and create topological defects (black dot), reducing the coherence length and amplitude of the CO order parameter.

# Supplementary Information

## Materials and Methods

*Materials:*
Single crystals of $La_{1.75}Sr_{0.25}NiO_4$ were grown using the floating zone method. The crystals were first oriented using Laue diffraction and cut along the (1, 0, 0) surface. Orthorhombic notation was used to express the crystal structure were [1, 0, 0] is 45 degree to the Ni-O bond direction. The surface was then polished for experiments. The phase transition temperature for charge order is approximately 110 K.

*X-ray FEL Configurations:*
Experiments were performed in the high charge mode (250 pC, 70 fs) of the LCLS at a repetition rate of 60 Hz. X-ray absorption spectrum (XAS) near the Ni $L_3$-edge was first measured to calibrate the incident photon energy (Fig. S1a). Then, the incident photon energy was tuned to the Ni-$L3$ peak in XAS to perform resonant diffraction on SO and CO. The scattering geometry is sketched in Fig. S1b. The chosen band width of approximately 1.6 eV covers the energy range of the primary resonance of the CO/SO diffraction peaks[15], yielding the maximal intensity. The transverse beam spot on the sample was 300 μm x 300 μm and the front-end N2 gas attenuator (set at approximately 10-30% transmission) was used to avoid radiation damage from the LCLS beam. The average FEL pulse energy was 0.5 mJ/pulse. With an estimated photon band pass associated with the selected resolution (16%) and the combined beamline efficiency (2.5%), we estimated the FEL fluence on sample surface to be approximately 0.6 mJ/cm$^2$. The polarization of the FEL is in the scattering plane.

*Pump-Probe Configurations:*
Photoexcitation was achieved via laser pulses with central wavelength at 800 nm from a Ti:sapphire laser amplifier system. Spatial overlap between the pump laser and x-ray FEL was achieved using the x-ray induced fluorescence of an unpolished YAG crystal, viewed by three cameras from three different angles. To ensure a homogeneous excitation of the region probed by the x-ray FEL, the pump laser spot diameter (FWHM) was set at 720 μm, which is more than twice as large as the x-ray FEL spot. The pump fluence was varied with a combination of the neutral density filters and a wave-plate and polarizer. The excitation density was estimated by taking into account the projected area on the sample surface corresponding to the angle of incidence, the absorption coefficient determined by reflectivity measurements, and the optical penetration depth calculated from the measured complex index of refraction on the same sample (see next section). The direction of pump pulse polarization is the same as the x-ray FEL pulse, in the scattering plane.

Coarse temporal overlap between the optical laser-pump and the FEL-probe pulses was achieved by monitoring the photoelectric signal of the two beams when overlapped onto the tip of a coaxial cable inside the vacuum chamber. The signal from this coaxial cable was measured directly with a 13GHz bandwidth oscilloscope. By matching the rising edges of the time traces from the two signals, the temporal overlap could be determined with < 3ps accuracy. Once coarse timing was established, the pump-probe delay time was varied with a mechanical delay stage. The fine temporal overlap was determined directly by monitoring the sample diffraction signal vs time-delay. The time zero was defined to be the onset of pump-probe signal. The data were taken by randomly sampling the delay time within a given range.

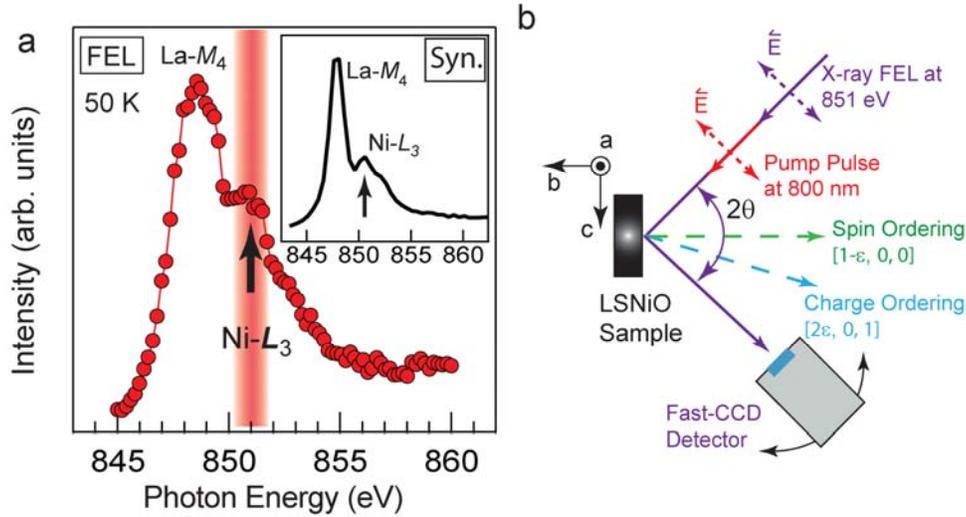

**Figure S1** (a) X-ray absorption spectrum (XAS) near Ni-$L_3$ edge measured by sample-to-ground current (i.e. the total electron yield) using the x-ray FEL with a monochromator. The red shaded area indicates the photon energy and bandwidth (1.6 eV) for our measurements. Each data point is an average over approximately 1000 x-ray FEL pulses. Inset shows the XAS obtained at the Advanced Light Source (ALS), a 3$^{rd}$ synchrotron light source, with a much better energy resolution (0.2 eV). (b) Scattering geometry of the experiment. The polarization of both x-ray FEL and pump laser pulses lies in the scattering plane (π-polarization). X-ray FEL and pump laser beams propagate co-linearly. The scattering angles (2θ) for the SO and CO are 156° and 143° degree, respectively. The incommensurability ε is found to be approximately 0.277. Orthorhombic notation [14] is used to express the position of the SO and CO in the reciprocal space.

*Data Collection and Process:*

A compact fast-CCD (cFCCD), developed by the Engineering Division of Lawrence Berkeley Laboratory[19], was used to record a two dimensional cross-section of the diffraction peaks (see Fig. S1c). It was mounted on an in-vacuum stage to allow the movement of the detector to cover almost the entire lower scattering hemisphere to image the diffraction peaks. The cFCCD has a maximum readout rate of 200 frames per second, allowing data acquisition on a pulse-by-pulse basis. The active area has a 240 (V) by 480 (H) array of 30 μm square pixels. At 100 mm away from sample, the detector pixel resolution was better than 0.02° and the effective acceptance angle was around 4° (V) by 8° (H).

An image was collected for every LCLS pulse along with the encoder position of the mechanical delay stage and corresponding x-ray pulse information for that image. In particular, information of phase cavity that measures the arriving time of the electron bunch was also recorded to partially correct timing jitter between the pump laser and x-ray pulse[21]. After each data run, approximately one thousand images were collected with the x-ray shutter closed and averaged together as a representative dark count image. This dark count image was subtracted from every image recorded in the associated data run. Due to small thermal drifts and fluctuations in the dark counts with time, an additional background adjustment was implemented for each shot using a small integrated region at the corner of the CCD image. The pump-probe delay time was calculated from the encoder position and phase cavity information. Then the images were binned and averaged according to the measured delay time. In addition, each image was also normalized by the incident photon flux ($I_0$) which was measured through the photocurrent from an in-line aluminum foil. Due to the LCLS energy and intensity fluctuations, some of the images contained very weak signals due to low x-ray intensity while still having similar thermal noise inherent to the CCD array. Thus an $I_0$ threshold was used to eliminate the images from the weaker LCLS pulses to maximize the signal-to-noise ratio. The $I_0$ threshold was set to eliminate approximately half of the collected images.

The cFCCD has inherent gain and offset variations amongst the different active arrays, causing the stripe-like feature on the collected images. After the images were normalized, binned and averaged, a ten pixel wide region at the edge of the image, away from the peak, was used to generate an instrument transmission function. The image intensities were then adjusted according to this transmission function and were processed through a low pass filter to yield a smoother image. A region of interest was chosen to measure the integrated peak intensity (white boxes in Fig. 1c of the main text) from the processed average image representing the different delay times. We note that the integrated intensity is rather insensitive to the image processing we have applied.

### *Optical Pump-probe Reflectivity Measurement:*
The ultrafast optical reflectivity change was measured by a pump-probe experimental setup based on a femtosecond Ti:Sapphire laser system. Ultrafast laser pulses at kilo-hertz repetition rate can be generated from this system with 100 fs pulse duration, 800 nm center wavelength, and 0.5 mJ per pulse energy. The pulses were split into pump and probe arm and then recombined onto the nickelate sample surface. The probe pulse reflected from the sample was detected by a photo diode detector and its reflectivity change due to the pump pulse was measured by a standard lock-in detection scheme. The sample was kept in a closed-cycle helium cryostat, where the temperature can be controlled from room temperature to 20K. The measurement geometry is similar to Fig. S1b, except the angle of incident is approximately 5 degree. The polarization of the pump is in the plane of incidence (b-c plane of the nickelate), while the polarization of probe is perpendicular to the plane of incidence (a-b plane of the nickelate).

## Estimation of the Excitation Density

The excitation density $\rho(\theta_{AOI})$ is estimated by

$$\rho(\theta_{AOI}) = F_{abs}/\delta(\theta_{AOI}),$$

Where $F_{abs}$ is the absorbed fluence, and $\delta(\theta_{AOI})$ is the intensity penetration depth of the pump pulse.

In order to calculate $F_{abs}$ and $\delta$ as function of $\theta_{AOI}$ we have characterized the complex refractive indices $n_{ab}$ and $n_c$ of $La_{1.75}Sr_{0.25}NiO_4$ along the a/b and c-axis. The reflectivity $R$ as function of $\theta_{AOI}$ was measured employing a modelocked Ti:sapphire oscillator system operating at the same central wavelength of 800 nm used in pump-probe time-resolved diffraction experiments. The incident beam is focused to ~ 100 μm spot size resulting in an incident fluences of 16 μJ/cm$^2$ at normal incidence. We recorded the reflectivity as function of $\theta_{AOI}$ and polarization for a plane of incidence being parallel to the b- and c-axis, respectively.

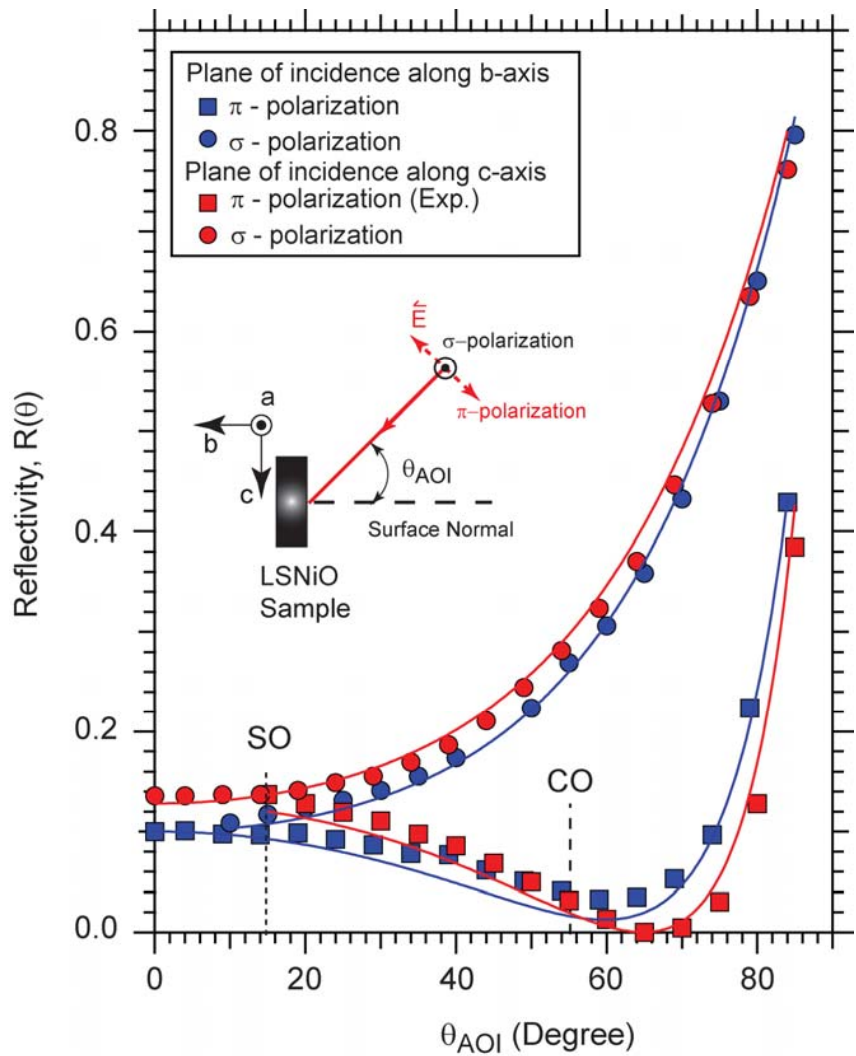

**Figure S2 Reflectivity of $La_{1.75}Sr_{0.25}NiO_4$ measured at a wavelength of 800 nm. The solid lines represent a global fit of the Fresnel formula for a uniaxial birefringent material [3]. Together with the optical conductivity along the a/b- and c-axis, a global fit yields complex refractive indices of $n_{ab} = 1.677 + i\, 0.536$ and of $n_c = 2.102 + i\, 0.143$.**

Fig. S2 shows the measured reflectivity $R$ as a function of $\theta_{AOI}$ for all four investigated high-symmetry geometries. The solid lines depict the results of a global fit of a set of Fresnel formulas for a uniaxially anisotropic material[3]. The imaginary part of the refractive index, deduced from optical conductivity data and the real part of the refractive index, has been used as an additional constraint in the global fitting process. According to the literatures[1,2], the optical conductivity for the doping level corresponding to our sample are $\sigma_{ab}$ = 3.75 x 10$^4$ Ω$^{-1}$m$^{-1}$ and $\sigma_b$ = 1.25 x 10$^4$ Ω$^{-1}$m$^{-1}$ along a/b- and c-axis, respectively. We obtained the complex refractive indices of $n_{ab}$ = 1.677 + $i$ 0.536 and of $n_c$ = 2.102 + $i$ 0.143. The error bars are estimated to be ±10%.

For the same geometry used for the pump-probe experiment (plane of incidence along c-axis, p-polarization), we observe that $R$ is approximately 4% for $\theta_{AOI}$ = 55° (CO geometry). Together with the projected beam footprint on sample, the absorbed fluence can be evaluated. Assuming that the absorption process is linear, the optical intensity penetration depth $\delta$ for the diffraction geometry of CO is calculated to be 260 nm, using the index ellipsoid of a uniaxial material with estimated $n_{ab}$ and $n_c$. For the optical pump-probe experiment, the $\theta_{AOI}$ is approximately 5°, where $R$ is approximately 15% and the penetration depth is estimated to be 440 nm.

## Estimation of x-ray Probe Depth and its Comparison to the Optical Intensity Penetration depth

It is also important to estimate the probe depth of the x-rays and compare that to the penetration depth of the optical pump laser. The x-ray penetration depth was estimated using the web tool provided on the website of the Center of X-ray Optics (CXRO), Lawrence Berkeley Lab (http://henke.lbl.gov/optical_constants/atten2.html). We used $La_{1.75}Sr_{0.25}NiO_4$ as the chemical formula, and calculate the angle dependence of the penetration depth at the Ni $L_3$-edge. It is found that the x-ray penetration depth at the Ni $L_3$-edge is approximately 80 nm for the scattering geometries used for measuring CO. Since the diffraction measurements were performed using reflection geometry, the x-ray probe depth can be estimated as the half the penetration depth. Thus, the probe depth of the x-ray pulse at Ni $L_3$-edge is estimated to be 40 nm for the scattering geometries of CO. We note that the x-ray probe depth is significantly smaller than the penetration depth of pump laser pulse for CO scattering geometries. Therefore, the x-ray probe volume is significantly smaller than the pump excited volume. This suggests that the spatiotemporal transport effect could be neglected in our discussion.

## Two Time Scale Model for Fitting

The CO resonant diffraction peak intensity could be written as

$$I_{CO} = |\Delta|^2 e^{-w_{phase}},$$

Where $\Delta$ is the order parameter's amplitude and $w_{phase}$ is the Debye-Waller factor due to order parameter's phase excitations. Therefore, we assume the recovery of the CO intensity as the product of two exponentially recovery processes corresponding to the amplitude and phase restorations, respectively. Namely, the time dependent CO normalized diffraction peak intensity can be expressed as:

$$I_{CO}(t) \sim (1 - A_a \cdot e^{-t/\tau_a}) \cdot (1 - A_p \cdot e^{-t/\tau_p}) = 1 - A_a e^{-t/\tau_a} - A_p \cdot e^{-t/\tau_p} + A_a A_p e^{-t(1/\tau_a + 1/\tau_p)}. \quad (1)$$

$A_a$ is the amount of suppression of the amplitude part (i.e. $|\Delta|^2$) of the CO peak intensity, while $A_p$ is the suppression of the phase part (i.e. $e^{-w_{phase}}$) of the CO peak intensity. $\tau_a$ and $\tau_a$ are the recovery time for CO peak intensity contributed from the order parameter's amplitude and phase, respectively.

We note that we have also fit the data to the commonly used two-time-scale model:

$$I_{CO}(t) \sim 1 - A_a e^{-t/\tau_a} - A_p \cdot e^{-t/\tau_p}. \quad (2)$$

The time scales obtained using this model are essentially identical to those fitted from the model first described. However, the $A_p$ and $A_p$ obtained are not accurate, since the product terms is not included in Eqn. (2).

For fitting the optical reflectivity data, the fit function is

$$R(t) \sim 1 - A_1 e^{-t/\tau_1} - A_2 \cdot e^{-t/\tau_2} + Const.$$

$A_1$ and $A_2$ are the magnitude of the suppression of the two dynamical processes, while $\tau_1$ and $\tau_2$ are their corresponding time scales. The constant term represents a long-living meta-stable state that is only observed in the optical data, but not in the resonant x-ray diffraction data. Thus, it is not directly related to the formation of long range CO. We note that this constant term can be either positive or negative, depending on the pump excitation density and temperature.

In the fitting process, time resolution is taken into account by a Gaussian convolution with corresponding FWHM, which is 0.4 ps and 0.1 ps for the resonant x-ray diffraction and optical reflectivity measurements, respectively.

**Supplementary References:**
1. Eisaki, H. & Uchida, S. Optical Study of High-Tc Superconductor and Related Oxides, *J. Phys. Chem. Solids* **56**, 1811-1814 (1995).
2. Ido, T., Magoshi, K., Eisaki, H. & Uchida, Optical study of the $La_{2-x}Sr_xNiO_4$ system: Effect of hole doping on the electronic structure of $NiO_2$ plane, *Phys. Rev. B* **44**, 12094 (1991).